\documentstyle[psfig]{mn}
\def\be{\begin{equation}}

\def\ee{\end{equation}}

\def\hcMpc{h^{-3}{\rm Mpc}^3}
\def\hcMpcinv{h^{3}{\rm Mpc}^{-3}}

\def\kms{\mbox{km s$^{-1}$}}

\newcommand{\phikb}{\phi(M_K, M_B)}
\newcommand{\Halpha}{\mbox{H$\alpha$}}
\newcommand{\ewh}{\mbox{EW\,(H$\alpha$)}}

\title[$K$-band galaxy luminosity function]
{The $K$-band luminosity function of nearby field galaxies}

\author[Jon Loveday]
{Jon Loveday\\
Department of Astronomy and Astrophysics, University of Chicago, 
5640 S. Ellis Ave, Chicago, IL 60637, USA\\
loveday@oddjob.uchicago.edu
}
\begin{document}

\maketitle

\begin{abstract}
We present a measurement of the $K$-band luminosity function (LF) of 
field galaxies obtained from near-infrared imaging of a sample of 345 galaxies 
selected from the Stromlo-APM Redshift Survey.
The LF is reasonably well-fit over the ten magnitude range $-26 \le M_K \le -16$
by a Schechter function with parameters $\alpha = -1.16 \pm 0.19$,
$M^* = -23.58 \pm 0.42$, $\phi^* = 0.012 \pm 0.008$ Mpc$^{-3}$,
assuming a Hubble constant of $H_0 = 100$ \kms Mpc$^{-1}$.
We have also estimated the LF for two subsets of galaxies subdivided by
the equivalent width of the \Halpha\ emission line at $\ewh = 10$\AA.
There is no significant difference in LF shape between the two samples,
although there is a hint  ($\sim 1 \sigma$ significance) that emission 
line galaxies (ELGs) have $M^*$ roughly one magnitude fainter than non-ELGs.
Contrary to the optical LF, there is no difference in faint-end slope $\alpha$
between the two samples.
\end{abstract}


\section{Introduction}

Deep, near-infrared $K$-band ($2.2\mu$m) galaxy surveys are a powerful tool for
studying galaxy evolution, eg. Gardner, Cowie \& Wainscoat \shortcite{gcw93},
Cowie et al.\ \shortcite{cghshw94}, Glazebrook et al.\ \shortcite{gpmc95}.
Compared to blue-optical light, near-infrared light is a better tracer
of mass in evolved stars and the correction for redshift
dimming (the ``$k$-correction'') is approximately independent of
morphological type.
The rapid evolution in galaxy number counts apparent in the $b_J$ band is not
seen in the $K$ band, eg. Koo \& Kron \shortcite{kk92}.
However, it is vital to have a reliable determination of the $K$-band
luminosity function (LF) for {\em nearby} galaxies in order to interpret
faint galaxy counts and to calculate the clustering of $K$-selected galaxy
samples.
Local $K$-band luminosity functions have been measured from optically-selected
samples by Mobasher, Sharples \& Ellis \shortcite{mse93} and 
Szokoly et al. \shortcite{sscm98}
and from $K$-selected samples by Glazebrook et al.\ \shortcite{gpmc95} and
Gardner et al. \shortcite{gsfc97}.
Since all of these surveys are flux-limited, the majority of galaxies have
$K$-band luminosities close to $L_K^*$.
Even in the largest sample (Gardner et al., 510 galaxies), it was
feasible to measure the LF over a range
of only 5 magnitudes, ${M^*}^{+3}_{-2}$, and so the the faint-end slope of 
the $K$-band luminosity function, so important for predicting galaxy number 
counts, is still rather poorly constrained.

In this paper we present a new estimate of the $K$-band LF over a range
of ten magnitudes, ${M^*}^{+8}_{-2}$, based on a subsample of galaxies
selected from the Stromlo-APM galaxy survey \cite{lpme96}.
This survey is an ideal source
for estimating the $K$-band LF since redshifts have already
been measured for 1797 galaxies with $b_J < 17.15$
over a very large volume of space.
The solid angle of the survey is 1.3 sr and the median redshift is
about 15,300 km/s.
The key to measuring the faint-end of the $K$-band LF without a huge
investment of telescope time is to observe galaxies selected by their
intrinsic luminosity rather than their apparent flux.
One can make use of the fact that near-infrared and optical
luminosities are correlated \cite{mes86,sci96}, 
in order to preferentially select galaxies of high and low luminosity and
thus sample the luminosity range more uniformly than a flux-limited sample.
One is thus able to measure the luminosity function to fainter
luminosities than from a flux-limited sample of similar size.
This sampling strategy is described in \S\ref{sec:sample} and
the observations and data reduction are discussed in \S\ref{sec:obs}.
Our method of estimating $\phi(M_K)$ from a $b_J$-selected sample is given
in \S\ref{sec:method} and we test this method in \S\ref{sec:test}.
Our results are presented in \S\ref{sec:results} and we conclude in
\S\ref{sec:concs}.
For notational convenience, we will denote absolute magnitudes 
in the $K$ and $b_J$ bands by $M_K$ and $M_B$ respectively.
Throughout, we assume a Hubble constant of $H_0 = 100$ km/s/Mpc.

\section{Sample Selection} \label{sec:sample}

The aim in selecting a subset of Stromlo-APM galaxies for which to obtain
$K$-band photometry was to sample the magnitude range 
$-22 \le M_B \le -13$ (the full range of $b_J$ absolute magnitudes
in the Stromlo-APM survey) as uniformly as possible.
An added complication in defining the sample arose because we wished
to obtain optical CCD images for the same sample of galaxies.
One planned use of this optical imaging is to measure morphological
parameters for a representative sample of galaxies at low redshift 
in order to compare with HST observations of galaxies
at high redshift, $z > 0.4$, \cite{brinch98}.
To obtain comparable linear resolution to the HST data required observing
galaxies at $z < 0.04$, assuming ground-based seeing of 1.3 arcsecond.
Our primary sample thus consists of galaxies at redshifts $z < 0.04$.
We divided the magnitude interval $-22 \le M_B \le -13$ into 90 bins
each of width 0.1 mag.
We then randomly selected up to six galaxies from the Stromlo-APM survey with
$z < 0.04$ in each bin.
Due to the redshift limit of $z < 0.04$, this primary sample
contains rather few galaxies brighter than $M_B = -20$.
We therefore formed a supplementary sample, consisting of galaxies at
$z > 0.04$ to top up each magnitude bin, where possible,
to six galaxies.
This supplementary sample consists entirely of galaxies with $M_B < -20$.
The primary sample contains 283 galaxies, and the supplementary sample
contains 80 galaxies, giving a total sample size of 363 galaxies.

\section{Observations and data reduction} \label{sec:obs}

Imaging of the above sample of galaxies was carried out at the
Cerro Tololo Interamerican Observatory (CTIO) 1.5m telescope using
the CIRIM infrared array through the standard $K$ filter
over the nine nights 1996 August 31 -- September 4 and 1997 October 19--22.
The pixel size at $f/7.5$ is $1.16''$, allowing most galaxies
to be observed at 9 non-overlapping positions on the $256 \times 256$ array.
Two frames were taken with the galaxy at the central position, 
thus yielding 10 frames per galaxy.
Total integration time for each galaxy was 300 seconds.
For 11 galaxies with angular size more than 100 arcsec,
we obtained four on-source and four offset-sky integrations of 75 seconds each.
In the following, these will be referred to as ``biggrid'' observations.
Standard stars were observed from the list of Elias et al.\ \shortcite{efmn82}.
Dark frames and dome flats were taken at the start of each night.

The infrafred frames were reduced using {\sc iraf}, 
mosaiced with the {\sc dimsum}\footnote{{\sc dimsum} is the Deep Infrared 
Mosaicing Software package developed by
Peter Eisenhardt, Mark Dickinson, Adam Stanford, and John Ward, and is
available via ftp from
ftp://iraf.noao.edu/iraf/contrib/dimsumV2/dimsum.tar.Z} package and 
image detection and photometry was performed using SExtractor \cite{ba96}.

\subsection{Basic reduction}

The basic reduction process consisted of the following steps:
\begin{enumerate}
\item Non-linearity correction using the {\sc irlincor} task with parameters
$c1 = 0.9997$, $c2 = 0.0257$, $c3 = 0.0158$ (Mike Keane, private communication),
\item Subtraction of dark frame,
\item Flatfielding by dome flat,
\item Masking of bad pixels.
\end{enumerate}

\subsection{Mosaicing}

At this stage of the reduction we had ten frames per galaxy, with the galaxy
at a different position on each frame.
Mosaicing of the frames and subtraction of sky background was performed
using the {\sc dimsum} package.
As described by Stanford, Eisenhardt \& Dickinson \shortcite{sed95},
{\sc dimsum} employs a two-pass procedure in order to mask out faint as
well as bright images when constructing the sky background, 
resulting in a much flatter sky than is obtained from median filtering of 
the galaxy frames with outlier rejection.
Individual galaxy frames were block replicated by a factor
of 4 in each dimension, allowing alignment to be performed by integer offsets
without interpolation.
Co-added images and corresponding exposure weight maps were made by
summing the aligned images and were finally block averaged $2 \times 2$ 
in order to economize on disc space.

For the 11 ``biggrid'' observations, the four offset sky frames were median 
filtered with outlier rejection in order to estimate the sky background.
The sky background was subtracted from each of the four on-source frames,
which were then aligned and coadded.

\subsection{Image detection and photometry}

We used SExtractor 2.0.15 to detect and measure images in the mosaiced frames.
For both standard stars and galaxies we used the {\sc mag\_best} estimate
of magnitude.
This yields a pseudo-total magnitude \cite{kron80} except in crowded fields,
when a corrected isophotal magnitude is measured instead.
Magnitude errors were estimated by combining in quadrature SExtractor's
estimate of the error from photon statistics and the difference between
magnitudes measured using local and global estimates of the sky background.
Of the selected sample of 363 galaxies, 351 were observed under photometric
conditions and 345 yielded a
$K$-band magnitude with an estimated error of less than 0.1 mag 
(rms mag error = 0.04 mag).

\subsection{Calibration}

Since we observed only single band infrared imaging, we used the following 
simple relation to convert observed magnitudes $k$ to standard CIT
\cite{efmn82} magnitudes $K$:
\be
K = k + k_0 + k_X X, \label{eqn:trans}
\ee
where $k_0$ is a zero-point offset, $k_X$ is the extinction coefficient in
the $K$-band and $X$ is the airmass of the observation.
We made a total of 74 standard star observations during the runs
(an average of 8 per night) and we initially
fitted the parameters $k_0$ and $k_X$ for all standard star observations
combined, obtaining $k_0 = 7.588 \pm 0.020$ and $k_X = 0.081 \pm 0.018$
with rms residual magnitude error of 0.018.
Six standard star observations with large residuals were omitted from the
fitting procedure, three of these were from the night of 1996 August 31,
which was partially non-photometric.
Holding the zero-point term fixed at $k_0 = 7.588$ we then fitted the
extinction coefficient separately for each night, with results shown
in Table~\ref{tab:calib}.
Galaxy magnitudes were converted to the CIT system using (\ref{eqn:trans})
with $k_0 = 7.588$ and $k_X$ coefficients from the Table.
Galaxies observed during non-photometric conditions on the nights of 
1996 August 31 and 1997 October 20 were rejected: these cases were obvious 
from the rapidly varying sky background.

\begin{table}
 \caption{Standard star calibration data.}
 \label{tab:calib}
 \begin{tabular}{lrrcc}
 \hline
 \hline
 Night & $N_{\rm std}^1$ & $N_{\rm del}^2$ & $k_X^3$ & rms\\
 \hline
1996 Aug 31 & 11 & 3 & $0.105 \pm 0.006$ & 0.017\\
1996 Sep 01 &  4 & 1 & $0.101 \pm 0.002$ & 0.003\\
1996 Sep 02 &  9 & 0 & $0.081 \pm 0.004$ & 0.014\\
1996 Sep 03 &  9 & 0 & $0.082 \pm 0.004$ & 0.015\\
1996 Sep 04 &  7 & 0 & $0.092 \pm 0.004$ & 0.011\\
1997 Oct 19 &  9 & 0 & $0.083 \pm 0.003$ & 0.010\\
1997 Oct 20 &  7 & 1 & $0.067 \pm 0.004$ & 0.010\\
1997 Oct 21 & 11 & 1 & $0.075 \pm 0.002$ & 0.007\\
1997 Oct 22 &  7 & 0 & $0.066 \pm 0.005$ & 0.013\\
  \hline
  \hline
 \end{tabular}

 \medskip
 $^1$ Number of standard stars observed.\\
 $^2$ Number of outliers deleted.\\
 $^3$ Extinction coefficient in (\ref{eqn:trans}).
\end{table}

\subsection{Photometric repeatability}

\begin{figure}
\centerline{\psfig{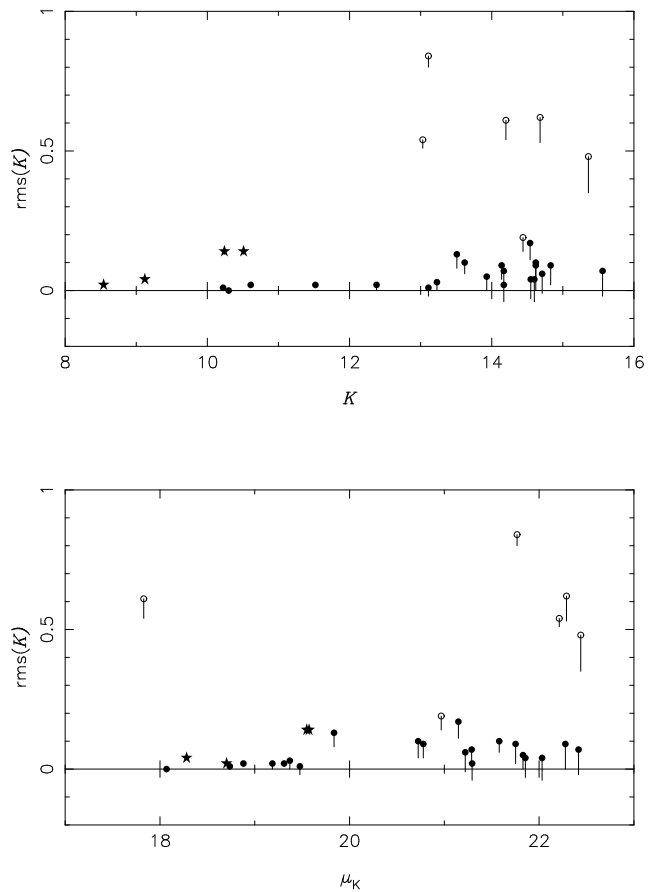}}
\caption{Repeatability of galaxy photometry.
The top panel plots the rms $K$ magnitude between repeated observations
against $K$ magnitude from the coadded data.
The symbols indicate the rms magnitude error and the length of
the line shows the estimated magnitude error.
The lower panel plots the same information but as a function of mean
$K$-band surface brightness $\mu_K$ within the measurement aperture.
Open symbols denote sources deblended by SExtractor, stars indicate a 
``biggrid'' observation.
\label{fig:repeats}}
\end{figure}

Thirty-one galaxies, mostly of low $K$-band surface brightness, were observed 
on more than one occasion, which allows us to assess the repeatability of
our photometry.
In these cases, we ran SExtractor on the mosaiced galaxy frames both before
and after coadding observations.
We obtain final photometry from the coadded frame and use 
the two individual frames and the coadded frame to estimate the rms error.
In four cases, one observation was made using ``biggrid'': in these cases
we do not coadd the observations and instead use the {\sc dimsum}-reduced 
observation for the final magnitude and estimated error.

In Figure~\ref{fig:repeats} we plot both our estimated errors and the rms
errors between repeated observations as a function of $K$ magnitude and mean
surface brightness $\mu_K$.
The symbols indicate the rms magnitude error and the length of
the line shows the estimated magnitude error from the coadded image.
Thus if our estimated errors are a good estimate of the true rms then the lower
ends of the error bars should reach zero.
In most cases, the rms error is within 2--3 times the estimated error 
and in a few cases the error is slightly overestimated.
The five points with rms $> 0.4$ all correspond to images which were broken
up by SExtractor's deblender (open symbols).
We attempted to sum the flux from the components, but clearly the
$K$ magnitude for these objects is accurate to only $\sim 0.5$ mag.
We see that most of the discrepant points are of extremely low surface 
brightness, $\mu_K \ga 21.5$ mag arcsec$^{-2}$.
For unbroken images, all rms errors are less than 0.2 mag and the estimated
errors provide a reasonable estimate of the rms.
The stars denote objects in which one of the observations used separate
sky exposures (``biggrid'' observations).
For two out of four of these objects, we see an rms magnitude error larger
than 0.1 but with a negligible estimated error (from the non-biggrid 
observation).
The reason for this is the poorer sky subtraction of the biggrid observations
compared with the {\sc dimsum} reduction of the majority of our observations.

\subsection{Matching to APM}

Since the mosaiced frames cover an area of sky around 8 arcmin on a side,
albeit not to uniform depth, they provide $K$-band photometry for
many objects in each field in addition to the target galaxy.
We therefore matched the images detected by SExtractor with images in the
APM scans.
Using the matched objects in each frame, we calculated a 6-parameter transform 
from CIRIM pixel coordinates to APM plate coordinates and thence to RA \& Dec.

The $K$-band photometry for the target galaxies is presented in the Appendix
to this paper.

\subsection{Sample properties}

\begin{figure}
\centerline{\psfig{figure=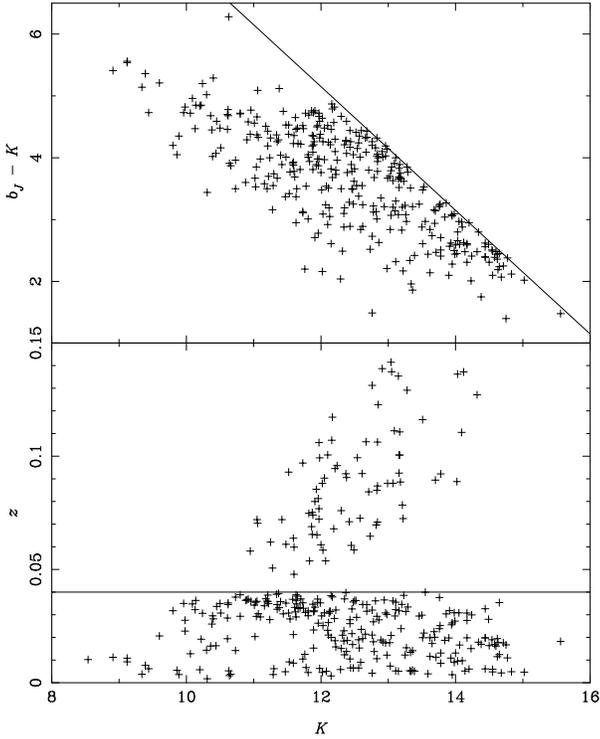,width=\linewidth}}
\caption{Illustration of observational selection effects in our sample.
The upper panel plots observed $b_j - K$ colour versus apparent $K$ magnitude.
The line shows the apparent magnitude limit for the Stromlo-APM survey, 
$b_J = 17.15$.
The lower panel plots redshift against apparent $K$ magnitude.
The line in this case separates the primary and supplementary samples
at $z = 0.04$.}
\label{fig:ksel}
\end{figure}

Since our source survey is limited by $b_J$ flux, our $K$-band sample is
not a complete one, as illustrated in Figure~\ref{fig:ksel}.
At faint $K$ magnitudes, only blue objects will be in the Stromlo-APM sample.
We are roughly complete to $K \approx 12$, the bluest galaxies can be seen
as faint as $K \approx 15$.
The luminosity function estimator described in \S\ref{sec:method} corrects
for this $K$-band incompleteness and allows us to use all galaxies with 
$K$-band photometry.
In the lower panel of this Figure we plot the redshift distribution as
a function of apparent $K$ magnitude.
The requirement that $z < 0.04$ for the primary sample has a noticeable
effect on the overall redshift distribution: this effect is also accounted
for by our luminosity function estimator.

\begin{figure}
\centerline{\psfig{figure=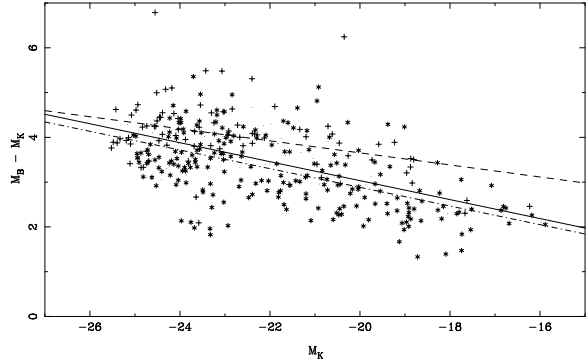,width=\linewidth,angle=-90}}
\caption{Rest-frame $M_B - M_K$ versus $M_k$ colour-magnitude plot.
	Plus signs represent early-type galaxies, asterisks late-type
	galaxies and dots represent unclassified galaxies.
	The solid line shows a least-squares fit to all galaxies, 
	the dashed line a fit to early types and the dot-dashed line a fit
	to late type galaxies.}
\label{fig:colmag}
\end{figure}

To calculate absolute magnitudes we use galactocentric recession velocities
and assume k-corrections in the $b_J$ band as given in Table~2 of
Efstathiou, Ellis \& Peterson \shortcite{eep88}.
We assume a $K$-band k-correction of $-2.5 z$ for all galaxy types
(ie. the near-infrared k-correction {\em brightens} galaxies with redshift).
This is a very good approximation to the $K$-band k-correction of
Glazebrook et al.\ \shortcite{gpmc95} for redshifts $z < 0.15$.

Figure~\ref{fig:colmag} shows the rest-frame $(M_B - M_K)$ versus $M_K$
colour-magnitude relation for our data.
The fit to all galaxies is given by
\begin{equation}
(M_B - M_K) = -0.212 \times M_K - 1.20,\ \sigma = 0.77. \label{eqn:colmag}
\end{equation}
The slopes in this relation for early and late-type galaxies
($-0.135 \pm 0.036$ and $-0.209 \pm 0.024$ respectively) are consistent
with those found for E-S0 ($-0.095 \pm 0.013$) and Sa-Sdm ($-0.24 \pm 0.03$)
galaxies by Mobasher, Ellis \& Sharples \shortcite{mes86}, although we find a
significantly larger scatter about the relation.
At least part of this scatter is due to saturated APM $b_J$ magnitudes 
for galaxies $b_J \la 15$ and a few instances of the APM scans breaking up a 
very large, bright galaxy into fragments.
An improved colour-magnitude relation will be available when we incorporate
optical CCD magnitudes for these galaxies (Loveday \& Lilly, in preparation).

\section{Estimating the $K$-Band Luminosity Function}
\label{sec:method}

When one has a sample selected on optical $b_J$ magnitude and wishes to 
estimate the $K$-band luminosity function,
the best way to proceed is to calculate a bivariate luminosity function 
(BLF, $\phikb$) allowing for known selection in $b_J$ flux and $M_B$ absolute
magnitude and then to integrate over $M_B$ to obtain $\phi(M_K)$.
One can estimate the shape of $\phikb$,
independently of inhomogeneities in the galaxy distribution, using
the maximum likelihood method of Sandage, Tammann \& Yahil \shortcite{sty79}.
The probability of seeing a galaxy with $K$-band luminosity $L_K^i$
and $B$ band luminosity $L_B^i$ at redshift $z_i$ in our sample is given by
\be
p_i = \frac{\phi(L_K^i, L_B^i) S(L_B^i)}
      {\int_{{L_K}_{\rm min}(z_i)}^{{L_K}_{\rm max}(z_i)}
      \int_{{L_B}_{\rm min}(z_i)}^{{L_B}_{\rm max}(z_i)}
      \phi(L_K, L_B) S(L_B) dL_K dL_B}.
\ee
The function $S(L_B)$ accounts for the known selection in absolute $B$ 
magnitude and the luminosity limits ${L_B}_{\rm min}(z_i)$ and 
${L_B}_{\rm max}(z_i)$
are the minimum and maximum $B$-band luminosities observable at
redshift $z_i$ in a sample limited by apparent $b$ magnitude.
For the sample analysed here, there are no flux limits in the $k$-band,
and so the integral over $K$-band luminosity runs from 
0 to $+\infty$.
The maximum-likelihood shape of the BLF $\phikb$ is estimated by maximizing
the likelihood ${\cal L} = \prod_{i=1}^{N_g} p_i$ (the product of the 
individual probabilities $p_i$ for the $N_g$ galaxies in the sample) 
with respect to the parameters describing the BLF.

\begin{figure}
\centerline{\psfig{figure=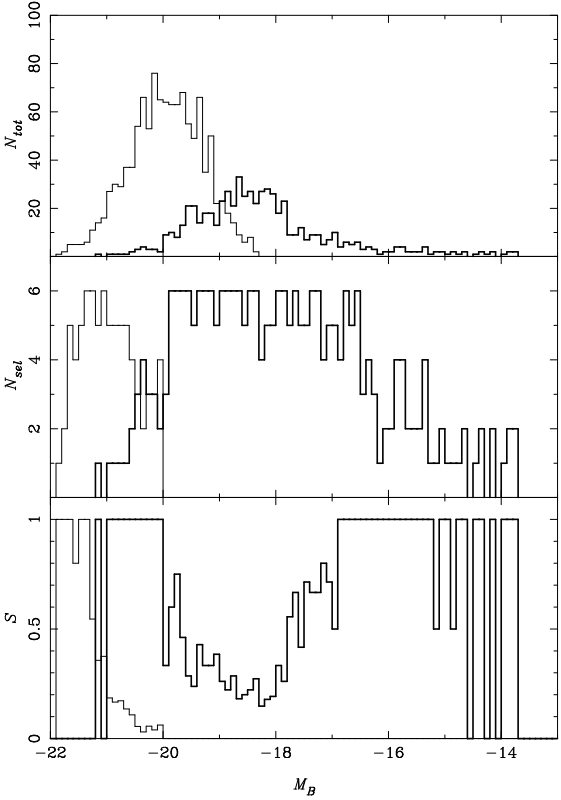,width=\linewidth}}
\caption{Sample selection as a function of $M_B$ absolute magnitude for the
primary (thick lines) and supplementary (thin lines) samples.
The upper panel shows histograms of $M_B$ for all 1797 galaxies in the
Stromlo-APM survey.
The middle panel shows the same for the 345 selected galaxies with $K$-band 
photometry.
The lower panel shows the sampling functions $S(M_B)$, the ratio of selected
to total galaxies in each bin.}
\label{fig:samp}
\end{figure}

In practice, we do not have a good {\em a priori} parametric model for 
$\phikb$, and so instead we measure $\phikb$\
in a non-parametric way using an extension of the Efstathiou, Ellis \& Peterson
\shortcite{eep88} stepwise maximum likelihood (SWML) method.
Sodr\'{e} \& Lahav \shortcite{sl93} have extended the SWML method to estimate
the bivariate diameter-luminosity function and to allow for sample 
incompleteness.
We adopt their extension of the SWML estimator here, including the sampling
function $S(M_B)$ separately for the primary and supplementary galaxy samples.
These sampling functions are illustrated in Figure~\ref{fig:samp}, which shows 
histograms of $M_B$ for the entire Stromlo-APM sample, for the selected 
galaxies with good $K$-band photometry and the sampling functions $S(M_B)$.

We measure the bivariate luminosity function in bins of absolute
$K$ and $B$ magnitude,
\be
\phikb = \phi_{jk},\ \ j = 1, \ldots, N_{M_K},\ \ k = 1, \ldots, N_{M_B},
\ee
where $M_K^j - \Delta M_K/2 < M_K < M_K^j + \Delta M_K/2$ and
$M_B^k - \Delta M_B/2 < M_B < M_B^k + \Delta M_B/2$.

The log-likelihood is given by
\begin{eqnarray}
\ln{\cal L} & = & \sum_{i=1}^{N_g} \sum_{j=1}^{N_{M_K}} \sum_{k=1}^{N_{M_B}}
	      W_{ijk} \ln[\phi_{jk} S(M_B^k)] - \nonumber \\
	      & & \sum_{i=1}^{N_g} \ln\left[
	      \sum_{j=1}^{N_{M_K}} \sum_{k=1}^{N_{M_B}}
	      H_{ijk} \phi_{jk} \right] + {\rm const}.
\label{eqn:logL}
\end{eqnarray}
Here $W_{ijk} \equiv W(M_K^i - M_K^j,\ M_B^i - M_B^k)$, with,
\be
W(x,y) = \left\{ \begin{array}{ll}
		1 & \mbox{if} -\Delta M_K/2 \le x \le \Delta M_K/2\\
		  & \mbox{and} -\Delta M_B/2 \le y \le \Delta M_B/2,\\
			0 & \mbox{otherwise.}
		    \end{array}
	    \right.
\ee
The sampling function $S(M_B)$ (Fig.~\ref{fig:samp})
is incorporated into the ramp function $H$.
Writing $M_K^- = M_K^j - \Delta M_K/2$,\ $M_K^+ = M_K^j + \Delta M_K/2$,\ 
$M_B^- = M_B^k - \Delta M_B/2$,\ and $M_B^+ = M_B^k + \Delta M_B/2$, then
\be
H_{ijk} = \frac{1}{\Delta M_K \Delta M_B} \int_{M_K^-}^{M_K'} dM_K
	  \int_{M_B^-}^{M_B'} S(M_B) dM_B,
\ee
where $M_K' = \max[M_K^-, \min(M_K^+, {M_K}_{\rm lim}^i)]$ and
$M_B' = \max[M_B^-, \min(M_B^+, {M_B}_{\rm lim}^i)]$.

To fix the otherwise arbitrary normalisation constant we apply the constraint
\be
g = \sum_j \sum_k \phi_{jk} \left(\frac{L_K}{L_f}\frac{L_B}{L_f}\right)^\beta
    \Delta M_K \Delta M_B - 1 = 0
\ee
with $\beta = 1.5$ and the fiducial luminosity $L_f$ corresponding to 
$M = -20$ using a Lagrangian multiplier $\lambda$.
The new likelihood $\ln {\cal L'} = \ln {\cal L} + \lambda g$ is maximised
with respect to the $\phi_{jk}$ and $\lambda$, requiring that $\lambda = 0$.

Setting $\partial\ln{\cal L'}/\partial\phi_{jk} = 0$, one arrives
at a maximum likelihood estimate for $\phi_{jk}$,
\be
\phi_{jk} = n_{jk}\left/\sum_i^{N_g}\left[
	    \frac{H_{ijk} \Delta M_K \Delta M_B}
	    {\left(\sum_{l=1}^{N_{M_K}} \sum_{m=1}^{N_{M_B}} \phi_{lm} H_{ilm} 
	    \Delta M_K \Delta M_B \right)} \right] \right.,
\label{eqn:phiKB}
\ee
where $n_{jk} = \sum_{i=1}^{N_g} W_{ijk}$ is the number of galaxies
in the $(j,k)th$ bin.
Errors in $\phi_{jk}$ are estimated via the inverse of the information matrix
(Efstathiou et al.\ 1988).

As with all density-independent estimators, information about the overall
normalisation is lost.
We therefore normalise our BLF to the mean density of galaxies with 
$-22 \le M_B \le -13$ in the full Stromlo-APM sample, 
$\bar{n} = 0.071 h^3{\rm Mpc}^{-3}$, calculated as described by Loveday et al.\
\shortcite{lpem92}.
(Note that the density $\bar{n} = 0.047 h^3{\rm Mpc}^{-3}$ quoted by 
Loveday et al.\ (1992) is for the restricted magnitude range 
$-22 \le M_B \le -15$.)

Finally, we obtain the $K$-band luminosity function by summing over
$B$ magnitude bins,
\be
\phi(M_K^j) = \sum_{k=1}^{N_{M_B}} \phi_{jk}.
\label{eqn:phiK}
\ee
One can then fit a given functional form, eg. a Schechter \shortcite{schec76}
function, to the stepwise estimate by least-squares.

\section{Test of the Method} \label{sec:test}

We have tested the above procedure by using it to
estimate the $K$-band luminosity function from a set of Monte Carlo 
simulations.
We generated nine mock Stromlo surveys by a Soneira \& Peebles \shortcite{sp78}
hierarchical clustering simulation with similar clustering properties to that
measured from the Stromlo-APM Survey \cite{lmep95}.
Each galaxy in the simulation was assigned a $K$-band luminosity 
drawn at random from
a Schechter function with $\alpha = -1.20$ and $M_K^* = -23.6$
and then assigned a $b_J$ magnitude according to our observed
colour-luminosity relation (\ref{eqn:colmag}).
Galaxies were selected on their apparent $b_J$ magnitude, $b_J < 17.15$.
This process was repeated until each simulation contained 2000 galaxies.
We then sampled each simulation by absolute $M_B$ magnitude as described
in \S\ref{sec:sample}, finally yielding an average of 359 galaxies
per simulation.
We calculated the K-band luminosity function $\phi(M_K)$ for each
simulation as described in \S\ref{sec:method} and fit a Schechter function
to each by least squares.
Averaging over the nine simulations, and estimating the BLF in bins of width
0.5 mag, we measure mean and rms Schechter
function parameters $\alpha = -1.17 \pm 0.07$, $M^* = -23.6 \pm 0.2$.
The errors on the mean values are $\sqrt 9$ times smaller than the quoted
rms scatter between the simulations.
Our estimates of $\alpha$ and $M^*$ are in excellent agreement with the
input luminosity function.

\begin{figure}
\centerline{\psfig{figure=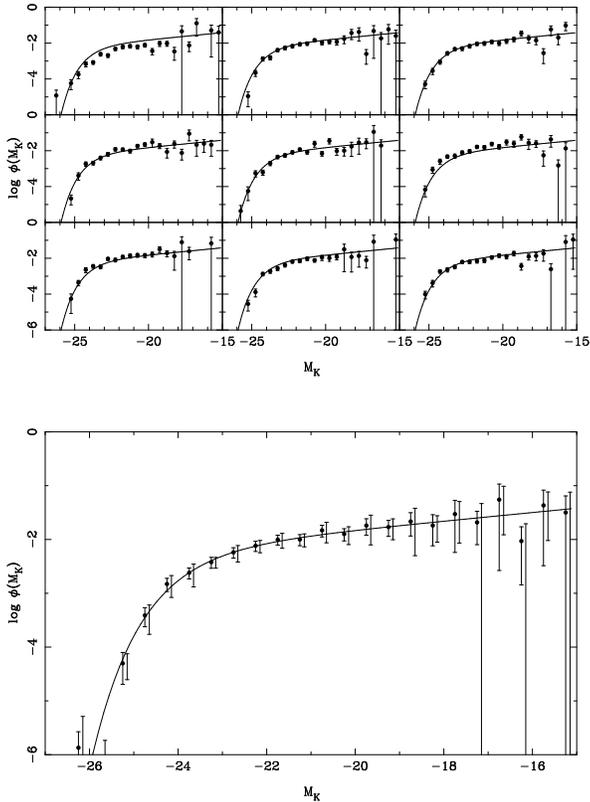,width=\linewidth}}
\caption{Test of our method to estimate the $K$-band luminosity function 
from $b_J$-limited samples.
The upper panels show the $K$-band LF estimated from each of nine 
Soneira-Peebles simulations and the lower panel shows the mean and rms
(error bars offset to right).
In each case the curve shows the {\em input} LF.
\label{fig:test}}
\end{figure}

The SWML estimates of the $K$-band LF for the simulations
are shown in Figure~\ref{fig:test}.
The points show the SWML estimate of $\phi(M_K)$ from
the nine simulations and the curve shows the input Schechter function with
shape $\alpha = -1.20$, $M^*_K = -23.6$.
The error bars going through the data points show the errors determined
from the covariance matrix.
The lower panel shows the average over the 9 simulations.
The rms scatter between realisations for each data point (shown by
the error bars offset slightly to the right) are in satisfactory agreement 
with the predicted errors.

Overall, we find that our procedure for estimating $\phi(M_K)$ from a
sample limited by apparent $b$ magnitude and further selected by absolute
$B$ magnitude provides a robust and unbiased estimate of the 
$K$-band LF over a wide range of absolute magnitudes.

\section{Results} \label{sec:results}

\begin{figure*}
\centerline{\psfig{figure=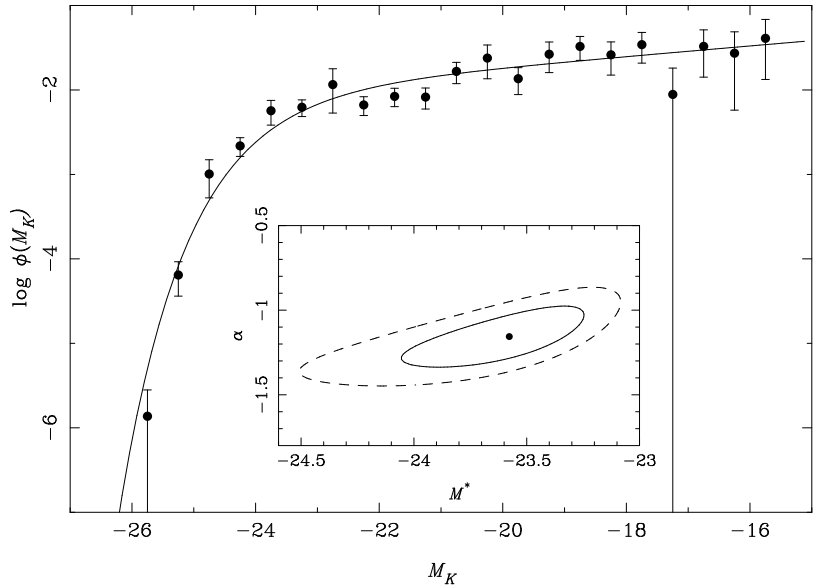,width=12cm,angle=-90}}
\caption{The $K$-band luminosity function estimated from our sample
(symbols) together with the best-fit Schechter function.
The inset shows the 1 and 2 $\sigma$ likelihood contours for the shape 
parameters $\alpha$ and $M_K^*$.}
\label{fig:klf}
\end{figure*}

Our estimated $K$-band luminosity function is shown in 
Figure~\ref{fig:klf}.
The curve shows a Schechter function fit to the SWML estimate using least
squares.
We allow for finite bin width by calculating the mean predicted 
$\langle \phi_j \rangle$ at the absolute magnitude of each galaxy in each bin, 
rather than simply calculating $\phi(M)$ at the bin centre.
The inset shows 1 and 2 $\sigma$ likelihood contours for the shape parameters
$\alpha$ and $M_K^*$, where the normalisation $\phi^*$ is adjusted to maximize 
the likelihood at each grid point.
The best-fit Schechter parameters are $\alpha = -1.16 \pm 0.19$, 
$M_K^* = -23.58 \pm 0.42$ and
$\phi^* = 0.0121  \pm 0.0082 h^3 {\rm Mpc}^{-3}$.
The quoted errors on $\alpha$ and $M_K^*$ come from the bounding-box of the 
1$\sigma$ likelihood contour.

Clearly the characteristic magnitude $M_K^*$ is rather poorly constrained
from these data, and so in order to measure the faint-end slope $\alpha$
independent of $M_K^*$, we have also fit a straight line to our estimated LF
over the restricted luminosity range $M > -22$.
We measure a slope $-1.08 \pm 0.15$, consistent with that measured
from the Schechter fit.
Furthermore, this slope does not change significantly varying the bright 
luminosity cut from $-20.5$ to $-23$.

Since we normalised the LF to the same number density as the full Stromlo-APM
sample, the error in $\phi^*$
is dominated by the uncertainty in shape of the LF.
The Schechter function provides a reasonable fit to the SWML estimate over the
full range of ten magnitudes; there is no indication of any faint-end turnup.

\begin{figure*}
\centerline{\psfig{figure=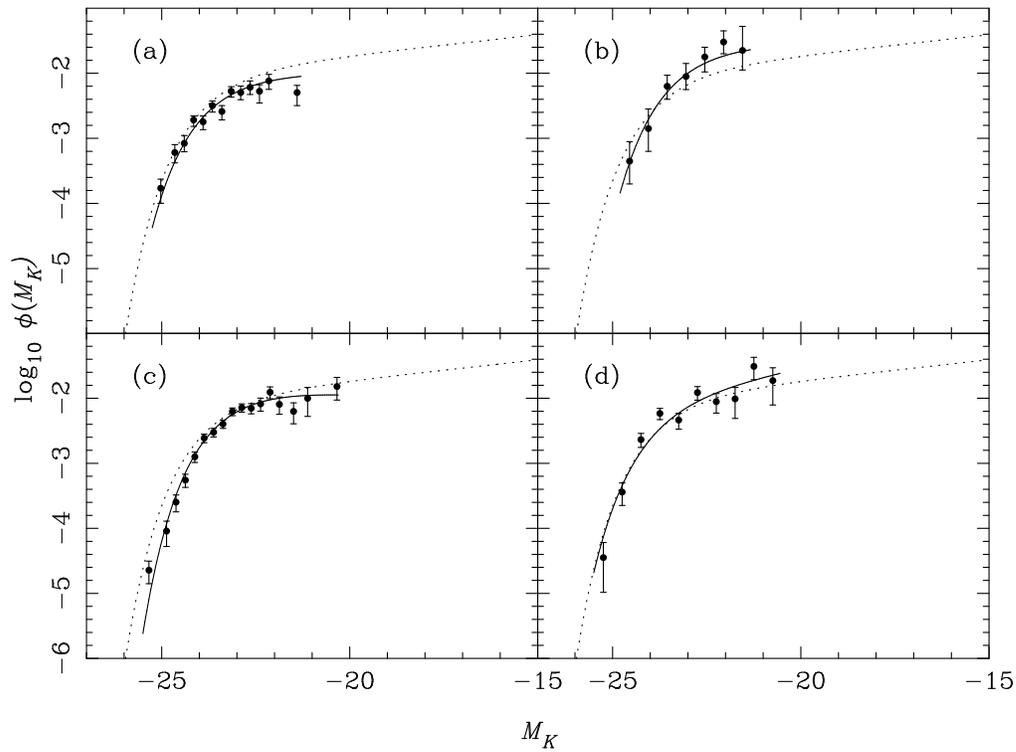,width=15cm,angle=-90}}
\caption{Comparison of our new estimate of the $K$-band luminosity function
(dotted line) with estimates from (a) Mobasher et al., (b) Glazebrook et al.,
(c) Gardner et al.\ and (d) Szokoly et al.}
\label{fig:comp}
\end{figure*}

\begin{table*}
\begin{minipage}{12cm}
 \begin{center}
 \caption{Schechter function fits to $K$-band LFs.}
 \label{tab:comp}
 \begin{math}
 \begin{array}{lccc}
 \hline
 \hline
 {\rm Sample} & \alpha & M^*_K & \phi^* / \hcMpc\\
 \hline
\mbox{Mobasher et al.\ 1993} & -1.0 \pm 0.3 & -23.4 \pm 0.3 & 0.0112 \pm 0.0016\\
\mbox{Glazebrook et al.\ 1995 ($z < 0.2$)} & -1.04 \pm 0.31 & -23.02 \pm 0.23 & 0.029 \pm 0.007\\
\mbox{Gardner et al.\ 1997} & -0.91 \pm 0.24 & -23.12 \pm 0.17 & 0.0166\\
\mbox{Szokoly et al.\ 1998} & -1.3 \pm 0.2 & -23.6 \pm 0.3 & 0.012 \pm 0.004\\
\mbox{This work (all galaxies)} & -1.16 \pm 0.19 & -23.58 \pm 0.42 & 0.012 \pm 0.008\\
\mbox{This work (non-ELG)} & -1.3 \pm 0.4 & -23.5 \pm 0.7 & 0.001 \pm 0.001\\
\mbox{This work (ELG)} & -1.2 \pm 0.4 & -22.5 \pm 0.8 & 0.011 \pm 0.013\\
  \hline
  \hline
 \end{array}
 \end{math}
\end{center}
\end{minipage}
\end{table*}

In Figure~\ref{fig:comp} and Table~\ref{tab:comp}, we compare our new estimate
of the $K$-band luminosity function with those of other workers.
Following Glazebrook et al.\ (1995) and Gardner et al.\ (1997), we have added 
0.22 mag to the points from Figure~2 of Mobasher et al.\ (1993) to account for
their method of calculating K-corrections.
We have also applied an aperture correction of $-0.30$ mag to the 
Glazebrook et al.\ data (see Gardner et al.\ 1997).
Ours is the first estimate of $\phi(M_K)$ fainter than $M_K = -20$.
Brighter than this all estimates are in reasonable agreement,
particularly given the uncertainties in the normalisation of the LF from
small samples of galaxies.
The normalisation of the Glazebrook et al.\ (1995) sample is about twice
that of all the other samples.
This is likely to be due to sampling fluctuations, since their survey covers
only 552 arcmin$^2$ and contains a total of only 124 galaxies.

Note that the bright-end points of our estimated LF have dropped significantly
since the preliminary analysis of Loveday \shortcite{love98car}.
This is largely due to the improved K-band photometry obtained by mosaiccing
the galaxy frames with {\sc dimsum} compared with a cruder algorithm used
earlier.

\subsection{ELG versus non-ELG}

As an aid to understanding the processes of galaxy formation and 
evolution, it is of great help 
to have estimates of the luminosity function for different types of galaxies.
The most objective way of separating Stromlo-APM galaxies is via the equivalent
width of the \Halpha\ emission line \cite[hereafter LTM]{ltm99}.
Due to the small size of our $K$-band sample, we subdivide into just two 
subsamples, those with $\ewh < 10$\AA\ (non-ELG, 138 galaxies) and those with 
$\ewh \ge 10$\AA\ (ELG, 134 galaxies).
For the remaining 71 galaxies, no \ewh\ measurement is available (LTM).
For both subsamples we recalculated the sampling function $S(M_B)$ appropriate
for galaxies with \ewh\ less or greater than 10\AA\ as appropriate.
The LFs were normalised to mean densities of $\bar{n} = 0.018\ \hcMpcinv$
(non-ELG) and $\bar{n} = 0.098\ \hcMpcinv$ (ELG) as determined for galaxies
with these emission line properties from the full Stromlo-APM sample.
The fact that the estimated mean density of ELGs is larger than that for
galaxies of any \ewh\ ($\bar{n} = 0.071 \hcMpcinv$, cf. \S\ref{sec:method})
is a consequence of the mean density of subsamples of Stromlo-APM galaxies
being determined to only $\sim 30\%$ accuracy (LTM).

\begin{figure}
\centerline{\psfig{figure=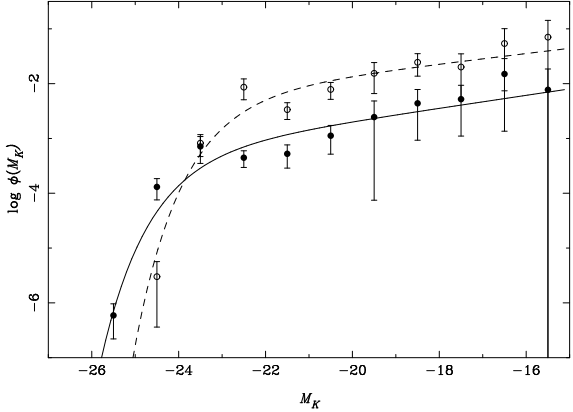,width=\linewidth,angle=-90}}
\caption{Comparison of the $K$-band luminosity function estimated from our 
non-ELG sample (filled symbols, continuous line) and ELG sample (open symbols,
dashed line).}
\label{fig:ew}
\end{figure}
\begin{figure}
\centerline{\psfig{figure=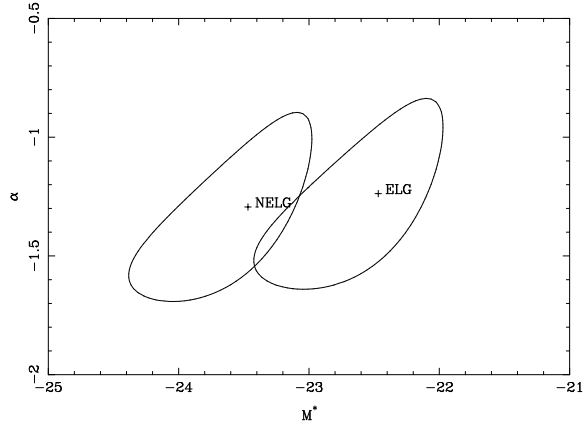,width=\linewidth,angle=-90}}
\caption{$1\sigma$ likelihood contours for the best fit Schechter parameters
for the non-ELG and ELG samples.}
\label{fig:ew_like}
\end{figure}

The estimated LFs for each subsample are shown in Figure~\ref{fig:ew}
and the $1\sigma$ likelihood contours for the best fit Schechter parameters
are shown in Figure~\ref{fig:ew_like}.
The Schechter function parameters themselves are listed in Table~\ref{tab:comp}.
Compared with the optical luminosity functions for \ewh-selected samples (LTM),
the $K$-band LFs for ELGs and non-ELGs are very similar in shape,
with just a $\sim 1\sigma$ indication that ELGs have characteristic magnitudes
$M^*$ about one magnitude fainter than non-ELGs.
Note that the faint-end slopes for the two samples are completely consistent.
These results are in accord with the LFs estimated for E/S0 and spiral 
galaxies by Mobasher et al.\ (1993).

\section{Conclusions} \label{sec:concs}

We have presented a new estimate of the $K$-band luminosity
function obtained from infrared array imaging of a subsample of galaxies
from the Stromlo-APM survey.
We measure $\phi(M_K)$ over a range of 10 magnitudes, ${M^*}^{+8}_{-2}$, 
a significantly greater range of luminosities than has been measured before now.
Our LF is consistent with earlier estimates, and shows that a
Schechter function of moderate faint-end slope ($\alpha = -1.16$) provides
a good fit to the LF as faint as $M^* + 8$ with no hint of any upturn
at the faint-end.
This observation appears to rule out the suggestion of Glazebrook et al.\ 
(1995) that there may be an extra dwarf component to the $K$-band LF, or
at least that if there is such a component, it must have characteristic 
magnitude $M_{\rm dwarf} \ga M^*_K + 8$.
In contrast, the faint-end upturn seen in the field $b_J$ LF \cite{love97}
is apparent by $M^*_B + 6$, and in the CfA Redshift Survey \cite{mhg94} by
$M^*_{Zw} + 3$.
In the Coma cluster, an upturn in the $H$-band
($1.6\mu m$) LF is also seen at $M^*_H + 3$ \cite{desd98}.
Note that we do see a slight upturn in $\phi(M_K)$ near $M^*_K + 3$,
faintward of the three low points around $M_K = -22$, but the LF flattens
out again by $M_K = -20$.

The $K$-band LFs for non-ELGs and ELGs are very similar in shape, with no
difference in faint-end slopes, contrary to what is seen in the optical LF 
(LTM).
The steep optical LF for ELGs might be consistent with a $\sim$ flat mass 
function in a fading burst model \cite{hp97}, 
although the particular model that Hogg \& Phinney consider
would steepen the $K$-band LF even more than the optical LF.

The main limitation with our sample is that the $b_J$ selection of the
source survey introduces a bias against red galaxies compared to a
$K$-selected sample.
However, the good agreement between the $K$-band LFs determined from $K$
and $b_J$-limited samples in Fig.~\ref{fig:comp} suggests that our
$b_J$ selection does not significantly affect the estimated $K$-band LF.
In particular, it is unlikely that we are underestimating the faint-end slope
of the $K$-band LF since red galaxies tend to be luminous in $K$
\cite{mes86,sci96}.
We will further address the issue of colour selection in a future paper 
in which we plan to parameterize the bivariate LF
$\phi(M_K,M_B)$ using CCD $b_J$ magnitudes in addition to the $K$-band data
presented here.

\section*{Acknowledgments}

It is a pleasure to thank George Efstathiou for suggesting this project,
the CTIO staff for their excellent support, and Bahram Mobasher, Jon Gardner,
Karl Glazebrook and Gyula Szokoly for sending me their $K$-band LF data points.

{}

\appendix

\section{The $K$-band data}

In Table~\ref{tab:the_cat} we present a sampling of our $K$-band photometric
data.
The complete catalogue will be available from the Astronomical Data Centre
(http://adc.gsfc.nasa.gov/).
The first five columns of this table come from the Stromlo-APM survey
(Loveday et al.\ 1996).
The subsequent seven columns are derived from the $K$-band data.
Each column in the table is described below.
\begin{description}

\item[(1) Name:]
Galaxy naming follows the same convention as the APM Bright Galaxy Catalogue
\cite{love96} and the Stromlo-APM Redshift Survey (Loveday et al.\ 1996).

\item[(2), (3) RA, dec:]
Right ascension (hours, minutes, seconds) and declination (degrees, arcminutes,
arcseconds) in 1950 coordinates.

\item[(4) $b_J$:]
$b_J$ magnitude.

\item[(5) $cz$:]
Heliocentric recession velocity in km/s.

\item[(6), (7) $K$, $K_{\rm err}$:]
$K$ magnitude and its estimated error.

\item[(8), (9) Maj, Min:]
Semi-major and minor axes of measurement ellipse in arcseconds.

\item[(10) PA:]
Position angle in degrees measured clockwise from south-north line.

\item[(11) Flags:]
Flags output by SExtractor \cite{ba96}.

\item[(12) Notes:\\]
\begin{enumerate}
\renewcommand{\theenumi}{\arabic{enumi}:}
\item Galaxy deblended by SExtractor; flux of components was summed.
\item ``Biggrid'' observation: photometry may be less reliable than most
galaxies.
\item NGC1672: APM deblended this galaxy.
\item Possibly non-photometric.
\item Central galaxy in close group of 3.
\item Double nucleus.
\end{enumerate}

\end{description}

\begin{table*}
\begin{minipage}{\textwidth}
 \caption{$K$-band galaxy photometry.}
 \label{tab:the_cat}
 \begin{tt}
 \begin{tabular}{rrrrrrrrrrrl}
  \hline
  \hline
   \multicolumn{1}{c}{\rm Name} & \multicolumn{1}{c}{\rm RA} & 
   \multicolumn{1}{c}{\rm Dec}
   & $b_J$ & $cz$ & $K$ & $K_{\rm err}$ & Maj & Min & PA & Flags & Notes\\
   \hline
   075+069-077 & 21 26 16.37 & -68 39 53.3 & 16.32 & 11033 & 12.50 &  0.02 &  16 &  11 &  46 &  3 \\
076-113-015 & 22 56 20.64 & -69 43 06.6 & 16.48 &  3813 & 13.25 &  0.04 &  32 &  14 &  61 &  2 \\
077+055+032 & 23 11 33.04 & -70 38 10.8 & 17.05 & 33310 & 14.09 &  0.05 &  13 &   8 & 104 &  3 \\
077+062-116 & 23 11 36.12 & -67 52 12.9 & 16.57 & 29923 & 11.98 &  0.02 &  19 &  15 & 172 &  3 \\
078-130-118 & 00 21 06.17 & -62 46 46.8 & 16.52 & 12128 & 13.55 &  0.04 &  12 &   7 & 116 &  3 & 1 \\
078+109+066 & 23 39 46.85 & -66 14 01.5 & 14.62 & 10150 & 11.91 &  0.07 &  44 &  20 &  86 &  2 & 2 \\
078+012+091 & 23 57 37.39 & -66 47 31.6 & 15.42 & 21754 & 11.05 &  0.01 &  25 &  23 &  69 & 18 \\
080-018-033 & 01 31 33.11 & -64 28 28.6 & 17.01 &  8088 & 13.95 &  0.05 &  16 &  12 & 136 &  0 \\
080+009+024 & 01 26 43.01 & -65 31 49.0 & 14.84 &  1624 & 11.47 &  0.01 &  25 &  22 &  12 &  0 \\
082+032-078 & 02 50 37.37 & -63 38 30.4 & 16.68 & 30356 & 13.16 &  0.03 &  15 &   7 & 110 &  2 \\
107-092-049 & 21 31 44.58 & -64 06 47.1 & 16.14 &  3179 &  8.54 &  0.00 &  86 &  43 & 174 & 48 \\
107-009+013 & 21 17 34.35 & -65 19 12.8 & 16.66 &  5118 & 14.06 &  0.07 &  18 &  16 & 175 &  2 \\
107-053-114 & 21 24 40.36 & -62 56 09.3 & 17.09 &  8507 & 14.49 &  0.08 &  19 &   8 & 152 &  0 \\
108+093+006 & 21 43 24.76 & -65 10 25.7 & 16.03 & 10497 & 13.20 &  0.04 &  18 &  14 &  24 &  0 \\
108-114+028 & 22 20 24.21 & -65 32 46.0 & 16.63 &  6146 & 13.69 &  0.05 &  26 &  14 &  89 &  0 \\
108-105+055 & 22 19 19.57 & -66 03 24.1 & 14.56 & 10779 & 11.03 &  0.01 &  47 &  21 &  47 & 16 \\
108-083-130 & 22 13 20.13 & -62 38 40.3 & 16.85 & 36934 & 12.85 &  0.03 &  14 &  13 & 123 &  0 \\
109+063+033 & 22 32 40.84 & -65 41 37.3 & 16.69 & 21798 & 11.97 &  0.02 &  18 &  14 & 130 &  0 \\
109+014+028 & 22 41 34.01 & -65 37 16.2 & 14.84 &  3269 & 11.73 &  0.02 &  26 &  25 &  31 &  0 \\
110+020+024 & 23 24 18.95 & -65 32 45.2 & 15.04 &  1991 & 11.30 &  0.02 &  82 &  38 & 150 & 18 \\
111+063-007 & 23 50 48.93 & -59 58 26.5 & 14.81 &  3320 & 13.05 &  0.54 &  53 &  25 &  31 &  0 & 2 \\
111+118+005 & 23 42 33.53 & -60 08 39.0 & 15.52 &  3414 & 14.24 &  0.24 &  30 &  18 &  73 &  2 & 2 \\
111+025+058 & 23 56 29.36 & -61 11 27.6 & 16.20 & 28872 & 12.24 &  0.02 &  24 &  15 & 179 &  0 \\
112+092-034 & 00 24 46.45 & -59 24 30.9 & 15.97 & 11668 & 12.15 &  0.02 &  18 &  12 &  75 &  0 \\
112-083-068 & 00 50 12.34 & -58 47 36.5 & 16.15 &  5150 & 14.75 &  0.08 &  13 &   9 & 180 &  0 \\
112+079+009 & 00 26 25.33 & -60 13 13.5 & 16.62 &  4720 & 11.89 &  0.02 &  38 &  27 &   3 & 16 \\
113+004+088 & 01 15 15.04 & -61 43 42.3 & 15.31 &  8591 & 11.54 &  0.01 &  30 &  11 &   7 & 16 \\
114-031+009 & 01 58 39.99 & -60 14 34.1 & 17.06 &  6743 & 13.53 &  0.04 &  16 &   6 & 123 &  0 \\
114-015+074 & 01 56 18.04 & -61 27 21.5 & 14.80 &  7002 &  9.98 &  0.01 &  49 &  39 & 108 & 16 \\
114-123-003 & 02 12 22.81 & -59 56 14.5 & 16.35 &  1471 & 15.60 &  0.14 &   9 &   5 &  93 &  0 \\
116-048-060 & 03 17 06.64 & -58 58 03.9 & 16.62 & 21466 & 12.84 &  0.03 &  11 &  11 &  68 &  0 \\
116-074+000 & 03 21 19.26 & -60 04 51.6 & 17.04 &  5664 & 15.56 &  0.09 &  16 &  11 &  85 &  3 \\
117+080-131 & 03 36 49.50 & -57 36 19.7 & 16.02 &  4952 & 12.22 &  0.02 &  23 &   9 &  31 &  0 \\
117+097-087 & 03 34 07.28 & -58 24 46.6 & 16.11 & 17774 & 12.49 &  0.02 &  16 &   8 &  96 &  0 \\
118-094-066 & 04 39 27.28 & -58 50 19.6 & 14.55 &  1176 & 10.64 &  0.01 &  26 &  25 &  36 & 16 \\
118-129-038 & 04 44 42.52 & -59 18 46.7 & 15.19 &  1278 &  7.49 &  0.00 & 108 &  54 &  90 & 16 & 3 \\
118-080+121 & 04 38 41.93 & -62 20 35.2 & 16.76 &  8488 & 14.17 &  0.06 &  16 &   9 &  72 &  3 & 1 \\
118-094+123 & 04 41 03.91 & -62 21 21.0 & 16.00 &  6157 & 13.90 &  0.05 &  21 &  18 &  60 &  0 \\
144+044-118 & 20 47 21.43 & -57 52 35.6 & 15.32 &  3233 & 12.43 &  0.02 &  21 &  17 & 155 &  2 \\
144-126-031 & 21 12 03.18 & -59 26 26.9 & 16.24 &  9493 & 11.89 &  0.02 &  11 &  11 & 144 &  0 \\
144+095-133 & 20 40 26.42 & -57 34 32.3 & 16.93 & 10964 & 12.52 &  0.02 &  16 &  10 & 128 &  0 \\
144+091+082 & 20 39 21.60 & -61 34 36.5 & 15.69 & 22306 & 11.88 &  0.02 &  22 &  15 &  88 &  2 \\
145+055+050 & 21 23 44.04 & -61 02 29.3 & 15.32 &  4404 & 10.30 &  0.01 &  34 &  12 &  82 &  2 \\
145-099-026 & 21 46 53.28 & -59 35 26.3 & 16.36 &  8053 & 12.11 &  0.02 &  19 &  14 &  40 &  2 \\
145+035+005 & 21 26 58.51 & -60 13 18.9 & 15.07 &  8660 & 10.61 &  0.01 &  43 &  32 & 136 & 18 \\
147-070+099 & 22 59 03.11 & -61 53 25.3 & 17.02 &  7787 & 13.94 &  0.05 &  23 &  20 &  85 &  2 \\
148+099-068 & 23 11 40.26 & -58 46 11.3 & 16.55 &  3376 & 13.93 &  0.05 &  16 &   8 &  30 &  0 \\
149-013-101 & 00 01 43.21 & -53 11 47.1 & 14.60 &  9773 & 10.13 &  0.01 &  32 &  25 & 169 & 16 \\
149+031+040 & 23 55 51.03 & -55 48 43.0 & 16.27 &  9477 & 11.62 &  0.01 &  22 &   7 &   9 &  0 \\

   \hline
  \end{tabular}
 \end{tt}
 \end{minipage}
\end{table*}
\end{document}